\documentclass{aa} 

\usepackage[authoryear]{natbib}
\bibpunct{(}{)}{;}{a}{}{,} 

\authorrunning{Orozco Su\'arez, Asensio Ramos, and Trujillo Bueno} 
\titlerunning{The magnetic field vector configuration of a solar prominence} 

\usepackage{graphicx}
\usepackage{txfonts}
\newcommand{\degree}{\ensuremath{^\circ}\/}

\begin{document}
   \title{The magnetic field configuration of a solar prominence inferred from spectropolarimetric observations in the \ion{He}{i}~10830~\AA~triplet}

   \author{D.\ Orozco Su\'arez \inst{1,2}, A.\ Asensio Ramos\inst{1,2}, 
   \and J.\ Trujillo Bueno\inst{1,2,3}}

  \institute{Instituto de Astrof\'isica de Canarias, E-38205 La Laguna, Tenerife, Spain, \email{dorozco@iac.es}
   \and 
 	Departamento de Astrof\'isica, Universidad de La Laguna, E-38206 La Laguna, Tenerife, Spain
           \and 
Consejo Superior de Investigaciones Cient\'ificas, Spain}

   \date{Received ; accepted }

\abstract
{The determination of the magnetic field vector in quiescent solar prominences is possible by interpreting the Hanle and Zeeman effects in spectral lines. However, observational measurements are scarce and lack high spatial resolution.}
{To determine the magnetic field vector configuration along a quiescent solar prominence by interpreting spectropolarimetric measurements in the \ion{He}{i}~1083.0~nm triplet obtained with the Tenerife Infrared Polarimeter installed at the German Vacuum Tower Telescope of the Observatorio del Teide.}
{The \ion{He}{i}~1083.0~nm triplet Stokes profiles are analyzed with an inversion code that takes into account the physics responsible of the polarization signals in this triplet. The results are put into a solar context with the help of extreme ultraviolet observations taken with the Solar Dynamic Observatory and the Solar Terrestrial Relations Observatory satellites.}
{For the most probable magnetic field vector configuration, the analysis depicts a mean field strength of 7~gauss. We do not find local variations in the field strength except that the field is, in average, lower in the prominence body than in the prominence feet, where the field strength reaches $\sim$~25~gauss. The averaged magnetic field inclination with respect to the local vertical is $\sim$~77\degr. The acute angle of the magnetic field vector with the prominence main axis is 24\degr\/ for the sinistral chirality case and 58\degr\/ for the dextral chirality. These inferences are in rough agreement with previous results obtained from the analysis of data acquired with lower spatial resolutions.} 
{}


 \maketitle

%

 \section{Introduction} 
  \label{sec:intro}
 
Although the first photographic plates of prominences were taken more than 150 years ago it took 100 years to discover, by means of the first spectropolarimetric measurements in prominences, that these solar structures are clear manifestations of the confinement of plasma within giant magnetic structures\footnote{There are many papers about solar prominences, but we recommend the reader to go first through the historical work of Einar Tandberg-Hanssen (e.g., \citealt{1998ASPC..150...11T,2011SoPh..269..237T}).}. Prominences, also referred to as filaments when observed against the solar disk, are cool, dense, magnetized formations of 10$^4$~K plasma embedded into the 10$^6$~K solar corona (for reviews see \citealt{2010SSRv..151..333M,2010SSRv..151..243L}). They are located above Polarity Inversion Lines (PILs or filament channels), i.e., the line that divides regions of opposite magnetic flux in the photosphere. Morphologically speaking, prominences can be separated in different classes \citep{1943ApJ....98....6P,1995ASSL..199.....T}. Among them, quiescent prominences are seen as large sheet like structures suspended above the solar surface and against gravity. The global structure of quiescent prominences change little with time, preserving their global shape during days and even weeks. Locally, they consist of fine and vertically oriented plasma structures, so-called threads, that evolve continually (e.g., \citealt{1976SoPh...49..283E,1994SoPh..150...81Z}). Recent observations taken with the Hinode satellite have revolutionized our knowledge of quiescent prominences fine scale structuring and dynamics; for instance, plasma oscillations, supersonic down-flows, or plasma instabilities like the Rayleigh-Taylor instability in prominence bubbles \citep{2008ApJ...676L..89B,2010ApJ...716.1288B,2008ApJ...689L..73C,2007Sci...318.1577O}.

The magnetic configuration of quiescent prominences has been investigated by using first the longitudinal Zeeman effect and later by measuring the full Stokes vector in spectral lines sensitive to the joint action of the Zeeman and Hanle effect (e.g., \citealt{1989ASSL..150...77L} and \citealt{2007ASPC..368..291L} for reviews). For instance, spectropolarimetric observations in the 
\ion{He}{I}~D3 multiplet at 587.6~nm have greatly contributed to the understanding of the magnetic field configuration in prominences \citep{1983SoPh...89....3A,1985SoPh...96..277Q,2003ApJ...598L..67C}. Full Stokes polarimetry in the \ion{He}{I}~D3 multiplet at 587.6~nm is accessible from several ground-based observatories, such as the French-Italian telescope (THEMIS) at the Observatorio del Teide, the Advanced Stokes Polarimeter at the Dunn Solar Telescope in Sacramento Peak, or the Istituto Ricerche Solari (IRSOL) observatory.  

Of particular interest to infer the magnetic field vector in prominences is the \ion{He}{I} triplet at 1083.0~nm. This spectral line can be clearly seen in emission in off-limb prominences (e.g., \citealt{2006ApJ...642..554M}) and in absorption in on-disk filaments (e.g., \citealt{1998ApJ...493..978L}). The \ion{He}{I} triplet is sensitive to the joint action of atomic level polarization (i.e., population imbalances and quantum coherence among the level's sublevels, generated by anisotropic radiation pumping) and the Hanle (modification of the atomic level polarization due to the presence of a magnetic field) and Zeeman effects \citep{2002Natur.415..403T,2007ApJ...655..642T}. This fact makes the \ion{He}{I}~1083.0~nm triplet sensitive to a wide range of field strengths from dG (Hanle) to kG (Zeeman). Importantly, the \ion{He}{I} 1083.0~nm triplet is easily observable with the Tenerife Infrared Polarimeter (TIP-II; \citealt{2007ASPC..368..611C}) installed at the German Vacuum Tower Telescope (VTT) of the Observatorio del Teide (Tenerife, Spain). Finally, an user-friendly diagnostic tool called ``HAZEL'' (from HAnle and ZEeman Light) is available for modeling and interpreting the \ion{He}{I}~1083.0~nm triplet polarization signals, easing the determination of the strength, inclination and azimuth of the magnetic field vector in many solar structures \citep{2008ApJ...683..542A}. The HAZEL code has already been used to analyze \ion{He}{I} 1083.0~nm triplet spectropolarimetric data of prominences \citep{2013hsa7.conf..786O}, spicules \citep{2010ApJ...708.1579C,2012ApJ...759...16M}, sunspot's super-penumbral fibrils \citep{2013ApJ...768..111S}, emerging flux regions \citep{2010MmSAI..81..625A}, and the quiet solar chromosphere \citep{2009ASPC..405..281A}. The HAZEL code has also been applied to \ion{He}{I}~D3 observations of prominences and spicules \citep{2011ASPC..437..109R}. 

However, the information we have about the spatial variations of the magnetic field vector in solar prominences is still very limited because of the insufficient spatial resolution of the observations, restricted to single point measurements (e.g., \citealt{1983SoPh...83..135L,1983SoPh...89....3A}), single slit measurements (e.g., \citealt{2006ApJ...642..554M}), or two-dimensional slit scans (e.g., \citealt{2003ApJ...598L..67C,2007ASPC..368..347M}) at spatial resolutions of about 2\arcsec, much lower than the sub-arcseconds resolutions achieved by the Hinode spacecraft in prominence broad-band imaging. We have an approximate picture of the global magnetic properties of quiescent solar prominences, mainly thanks to the information encoded in spectral lines sensitive to the Hanle and Zeeman effect. The magnetic field in quiescent prominences is rather uniform and has mean field strengths of tens of gauss, typically in the range of 3~G to 30~G. The magnetic field vector forms an acute angle of about 35\degr\/ with the prominence long axis \citep{1970SoPh...15..158T,1983SoPh...83..135L,1994SoPh..154..231B,2003ApJ...598L..67C}. The field lines are found to be highly inclined with respect to the local vertical (e.g., \citealt{1983SoPh...89....3A}). For instance, \cite{1983SoPh...83..135L} found a mean inclination of 60\degr\/ from the local vertical in a sampling of 15 prominences. Their data were limited to single point measurements and their estimated rms error was about 15\degr. More recently, \cite{2003ApJ...598L..67C,2005ApJ...622.1265C} inferred the vector field map in a quiescent prominence and found inclinations of about 90\degr\/ with respect to the local vertical. These authors also reported that the field can be organized in patches where it increases locally up to 80~G. The magnetic configuration seems to be different for polar crown prominences where the field is found to be inclined by about 25\degr\/ with respect to the solar radius vector through the observed point \citep{2006ApJ...642..554M}. Finally, it has been found that, for 75\% of the analyzed prominences, the perpendicular component of the magnetic field vector to the prominence long axis or PIL points to the opposite direction with respect to the photospheric magnetic field. In this case, they are classified as inverse polarity prominences \citep{1983SoPh...83..135L}. 

\begin{figure}[!t]
\begin{center}
\resizebox{\hsize}{!}{\includegraphics{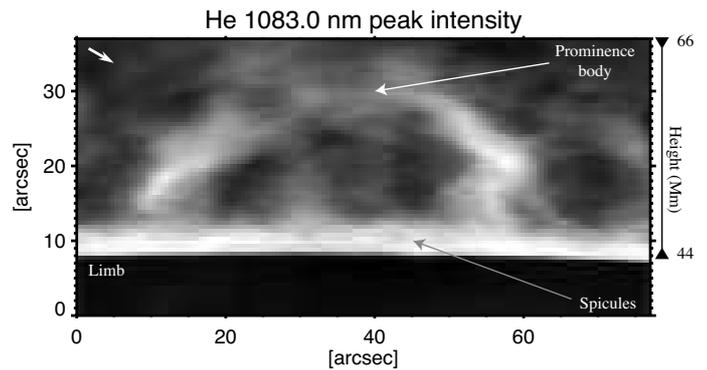}}
\end{center}
\caption{Peak intensity map of the \ion{He}{I}~1083.0~nm triplet emission profile. The prominence is seen as a bright structure against a dark background. The bottom, dark part corresponds to the solar limb. The top-right arrow points to the solar North direction. The de-projected height (see Sect.\ref{sec2})  above the solar surface is shown on the right axis. The data was taken on 20 May 2011, at 9:44 UT and finished at 11:15 UT, within the same day.}
\label{fig1}
\end{figure}

\begin{figure*}[!t]
\begin{center}
\resizebox{0.9\hsize}{!}{\includegraphics{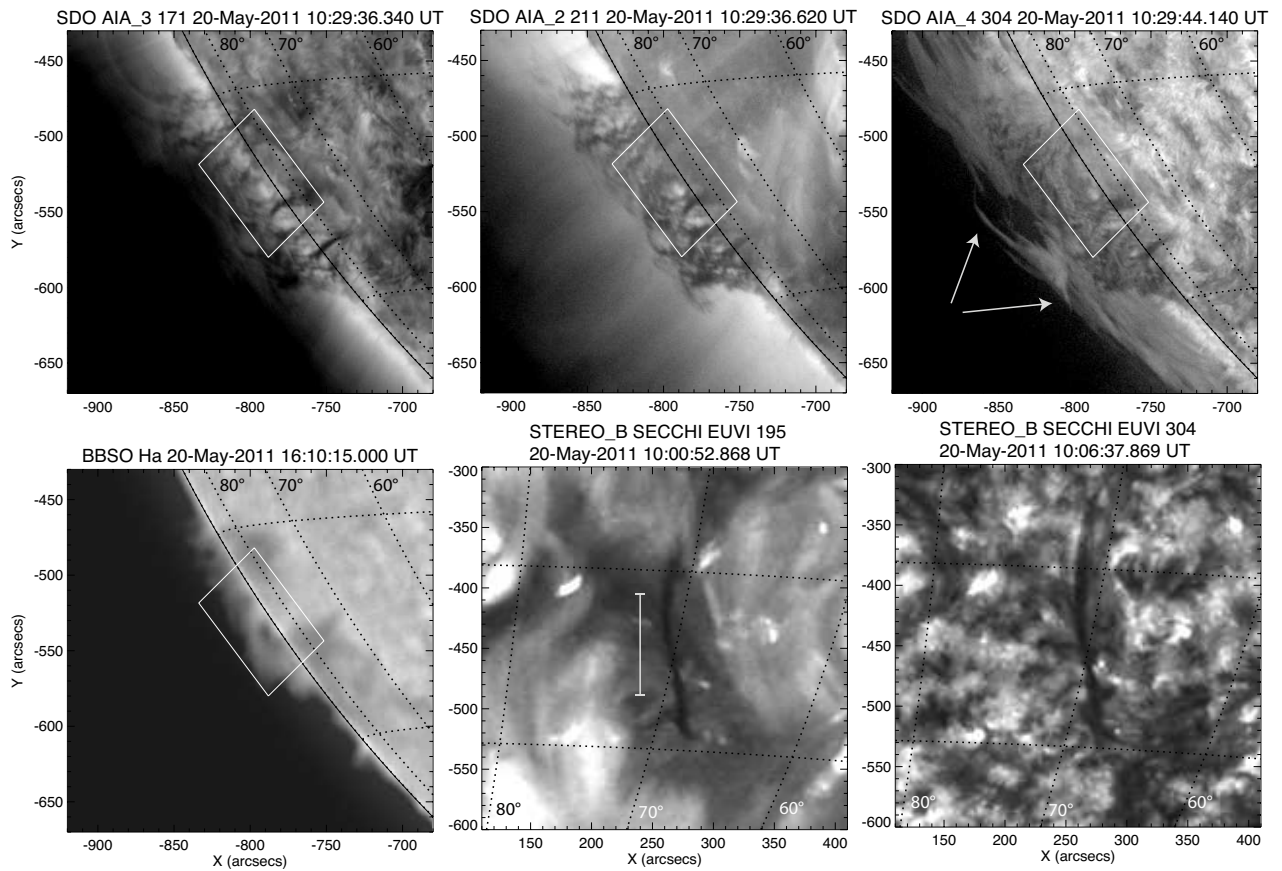}}
\end{center}
\caption{Illustrations of the observed prominence as seen in SDO/AIA, STEREO/EUVI, and the BBSO. Top panels correspond to \ion{Fe}{IX}~171~\AA,  \ion{Fe}{XIV}~211~\AA\/ and \ion{He}{II}~304~\AA\/ AIA band-pass filter images. Bottom panels are BBSO H$\alpha$ broad-band image, and \ion{Fe}{XII} 195 \AA\/ and \ion{He}{II} 304 \AA\/ STEREO-B/EUVI band-pass images. All images were taken on 20 May 2011, the same day the TIP-II observations were carried out. The white box represents the TIP-II field-of-view. Lines of constant Stonyhurst heliographic longitude and latitude on the solar disk are overplotted. Axis are in heliocentric coordinates. The arrows pinpoint the location of horn-like structures in \ion{He}{II} 304 \AA. The TIP-II slit virtual position can be seen in the botton central panel. The quiescent prominence can be clearly seen in the AIA images as well as in H$_\alpha$. In STEREO-B, it can be seen as a dark, elongated structure. The temporal evolution of the prominence in the SDO/AIA \ion{Fe}{XIV}~211~\AA\/ channel is available in the on-line edition.}
\label{fig2}
\end{figure*}

The magnetic field vector we infer through the interpretation of polarizations signals, such as those of the \ion{He}{I} 1083.0~nm multiplet, is associated with the coolest and densest prominence material. For this reason, the magnetic field in prominences has also been investigated by indirect means, i.e., constructing models of the field geometry in order to capture the observed prominence shape and properties \citep{1998A&A...329.1125A,1999A&A...342..867A,1998A&A...335..309A,2012ApJ...761....9D}. Such studies have contributed to our present picture of the global magnetic field structure associated to the prominence, although most models assume that the prominence material is suspended in magnetic dips. Among these models, we have the sheared-arcade models \citep{1989ApJ...343..971V} and the twisted flux-rope models \citep{1994SoPh..155...69R}. In the first ones, an helical magnetic structure is generated via photospheric shear flow motions that give rise to magnetic reconnection of pre-existing magnetic fields lines near the PIL. The cool material of the prominence is then supported by the magnetic dips of the helical structure via a magnetic tension force \citep{1957ZA.....43...36K,2005ApJ...626..551L,2010ApJ...714..618C}, by MHD-waves pressure \citep{2000SoPh..194...73P}, or by the presence of tangled magnetic fields in very small scales \citep{2010ApJ...711..164V}. The twisted flux-rope models suggest that the helical magnetic field structure supporting the prominence material has emerged from below the photosphere. Both models yield magnetic properties compatible with the present, low-resolution observational constraints. The local magnetic field in prominences has also been investigated by interpreting the dynamics of rising plumes using magnetohydrodynamic models \citep{2012ApJ...746..120H,2012ApJ...761..106H}.

Here we present the results of the analysis of ground-based spectropolarimetric observations of the \ion{He}{I}~1083.0~nm triplet taken in a quiescent solar prominence. The data were obtained with the TIP-II instrument \citep{2007ASPC..368..611C} installed at the German VTT at the Observatorio del Teide. This instrument is providing observations of solar prominences at spatial resolutions of about 1\arcsec--1\farcs5 during regular observing conditions, and even below one arcsecond in periods of  excellent seeing conditions. In this paper, some of these new observations will be analyzed with the HAZEL code. We will first describe the observations (sections 2 and 3) and then explain the diagnostic technique (section 4). In section 5 we present the inferred two dimensional map of the magnetic field vector of the observed prominence, and then we discuss and summarize the results in section 6.

\section{Observations, context data, and prominence morphology}
\label{sec2}

In this paper, we use observations taken with the TIP-II instrument on 20 May 2011. In particular, we scanned a region of about 40\arcsec\/ wide with the VTT spectrograph, crossing the solar south east limb where a quiescent prominence was visible in the H$_\alpha$ slit-jaw images. The observation started at 9:44 UT and finished at 11:15 UT. The length of the spectrograph's slit was 80\arcsec\/ with a spatial sampling along the slit of 0\farcs17 and a scanning step of 0\farcs5, which provided us with a 80\arcsec$\times$40\arcsec\/ map. Thanks to the adaptive optics system of the VTT we could maintain stable observing conditions (in terms of spatial resolution) during the 95 minutes that took the scanning of the prominence. We believe that the spatial resolution of our data lies between 1\arcsec\/ (limited by the scanning step) and 1\farcs5. During the scanning, the TIP-II instrument recorded the four Stokes parameters around the 1083.0~nm spectral region. This region contains the chromospheric \ion{He}{I}~1083.0~nm triplet as well as the photospheric \ion{Si}{I}~1082.70~nm line, including an atmospheric water vapor line at 1083.21~nm. The spectral sampling was 1.1~pm and the exposure time per polarization state was 15 seconds. The TIP-II data reduction process included dark current, flat-field, and fringes correction as well as the polarimetric calibration. To improve the signal to noise ratio the data were down-sampled spectrally and spatially along the slit direction, yielding a final spectral and spatial sampling of 4.4~pm and 0\farcs51, respectively. 

\begin{figure*}
\begin{center}
\resizebox{0.7\hsize}{!}{\includegraphics{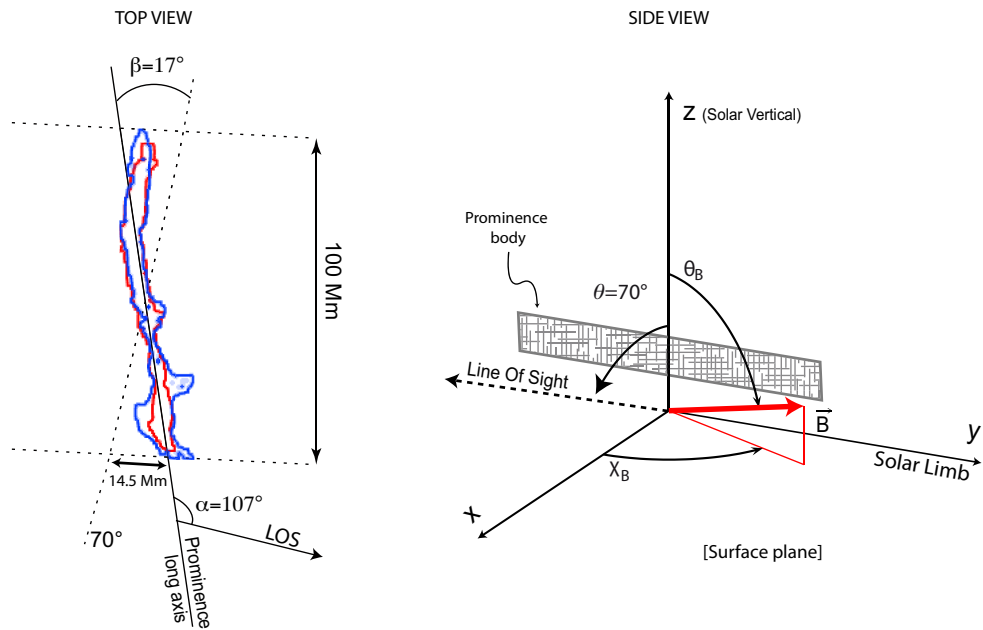}}
\end{center}
\caption{Left: Sketch of the prominence as seen on disk (top view). The blue and red contours outline the filament as seen in STEREO-B/EUVI \ion{Fe}{XII} 195 \AA\/ and \ion{He}{II} 304 \AA\/ band-pass images (see Figure~\ref{fig2}). The arrow points to the line-of-sight (LOS) direction. The dotted lines correspond to constant Stonyhurst heliographic longitude and latitude positions on the solar disk. The prominence is sitting in the 70\degr\/ longitude, what gives a scattering angle of $\theta = 70^\circ$. Right: Geometry of the problem. The inclination of the magnetic field  $\theta_\mathrm{B}$ is measured from the z-axis (solar vertical) and the azimuth $\chi_\mathrm{B}$ counterclockwise from the x-axis, contained in the surface plane. The graph shows the angle $\theta$ between the line-of-sight direction and the local vertical, which corresponds to the light scattering angle.}
\label{fig3}
\end{figure*}

The observed prominence can be seen in Fig.~\ref{fig1}, were the X-axis represents the position along the slit and the Y-axis is the scanning direction. The right axis shows the de-projected height over the solar surface. The more vertical appearance of the prominence at both sides of the observed FOV (hereafter, prominence feet) connected to each other with a more diffuse horizontal filamentary structure (hereafter, prominence body) shape the  prominence as loop-like structure. The feet show more \ion{He}{I} peak intensity signal that the prominence body. They may be connecting the prominence body with the chromosphere. In the intensity map, we cannot distinguish finer details within the prominence such as threads, despite the data were obtained during good seeing conditions. 

To put the TIP-II observations in context, we have made use of data provided by the extreme ultraviolet light telescope (AIA; \citealt{2012SoPh..275...17L}) onboard NASA's Solar Dynamics Observatory (SDO; \citealt{2012SoPh..275....3P}), the  Solar Terrestrial Relations Observatory (STEREO-B) Extreme UltraViolet Imager \citep{2008SSRv..136....5K}, and the Big Bear Solar Observatory (BBSO) high-resolution H$_\alpha$ filter \citep{1999SoPh..184...87D}. Figure~\ref{fig2} displays maps of the prominence as seen in the \ion{Fe}{IX}~171~\AA, \ion{Fe}{XIV}~211~\AA, and \ion{He}{II}~304~\AA\/ AIA band-passes (top panels). The AIA spatial resolution is about $\sim$1\farcs6. The white box outlines the TIP-II field-of view, thus our observations sampled the central part of the prominence as seen in these filters (see Fig.\ref{fig1}). In the 171~\AA\/ and 211~\AA\/ filters, the prominence is seen as a dark absorption against a bright background. It looks like a sheet made of strands, forming arc structures. In the 304~\AA\/ filter the prominence appearance is rather different. This filter shows radiation coming from plasma  at $\log T = 4.7$. In this case, it can be seen as a large and dark envelope with a horn-like (U-shaped) structure in the top (see arrows in Fig.~\ref{fig2}, top right panel). These ``horns'' may be indicating the presence of a coronal cavity right above the prominence \citep{2011A&A...533L...1R}.

\begin{figure*}[!t]
\begin{center}
\resizebox{\hsize}{!}{\includegraphics{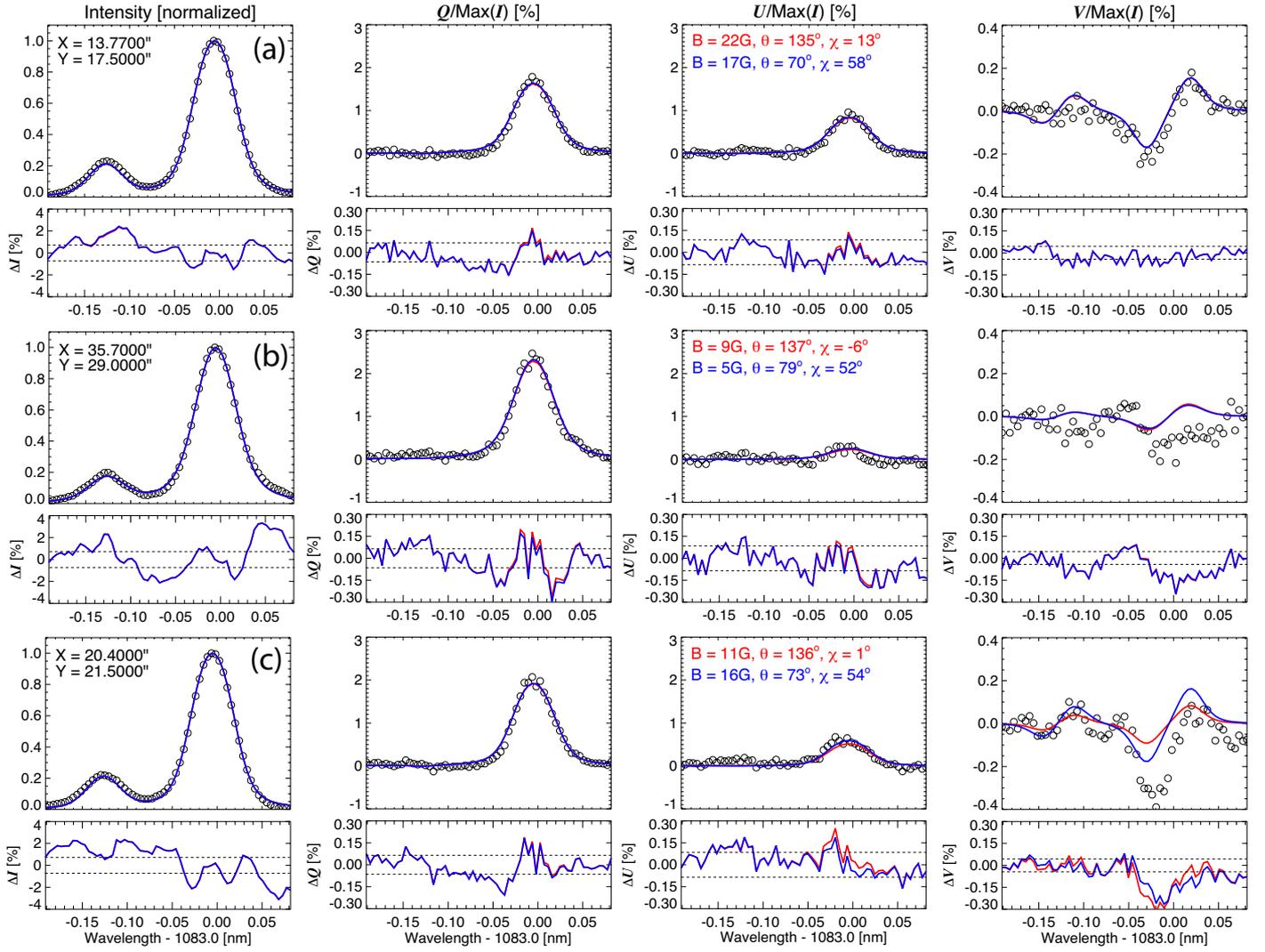}}
\end{center}
\caption{Observed and best-fit \ion{He}{1}~1083.0~nm triplet Stokes profiles 
corresponding to different pixels in the prominence. Stokes I is normalized to 
unity while Stokes Q, U, and V are normalized to their Stokes I maximum peak 
value. Open dots represent the observations and solid lines stand for the the 
theoretical profiles obtained by HAZEL. Blue and red color codes represent two 
different solutions. In this case, we plot the quasi-horizontal solution (blue) 
and the corresponding one to the 90\degr\/  ambiguity (red), which corresponds 
to the quasi-vertical solution (see Sect.~\ref{sec4}). The bottom sub-panels 
display the difference between the observed and the synthetic profiles. The 
horizontal dashed lines in the sub-panels stand for the mean standard deviation 
of the noise: $\overline{\sigma}_{I,Q,U,V} = \pm[0.72,0.065,0.085,0.045]$\%. They were  
calculated by averaging the standard deviation at a single wavelength point in 
the continuum and over all pixels showing polarization signal amplitudes above 
three times their corresponding noise level, in Stokes Q, U, or V. The legend 
in the Stokes I panels give the location of the pixel in arcseconds and that in 
Stokes U/I gives the values of the field strength, inclination, and azimuth 
retrieved for each pixel and for the two 90\degr\/ ambiguous solutions. The 
positive reference direction for Stokes Q is the parallel to the solar limb. 
Panels (a), (b), and (c) represent: a prototypical \ion{He}{1}~1083.0~nm triplet 
profile with significant linear and circular polarization signals, a profile 
with noisy Stokes V signal, and one whose Stokes V signal shows an anomalous 
profile shape. }
\label{fig4}
\end{figure*}

\begin{figure*}[!t]
\begin{center}
\resizebox{0.9\hsize}{!}{\includegraphics{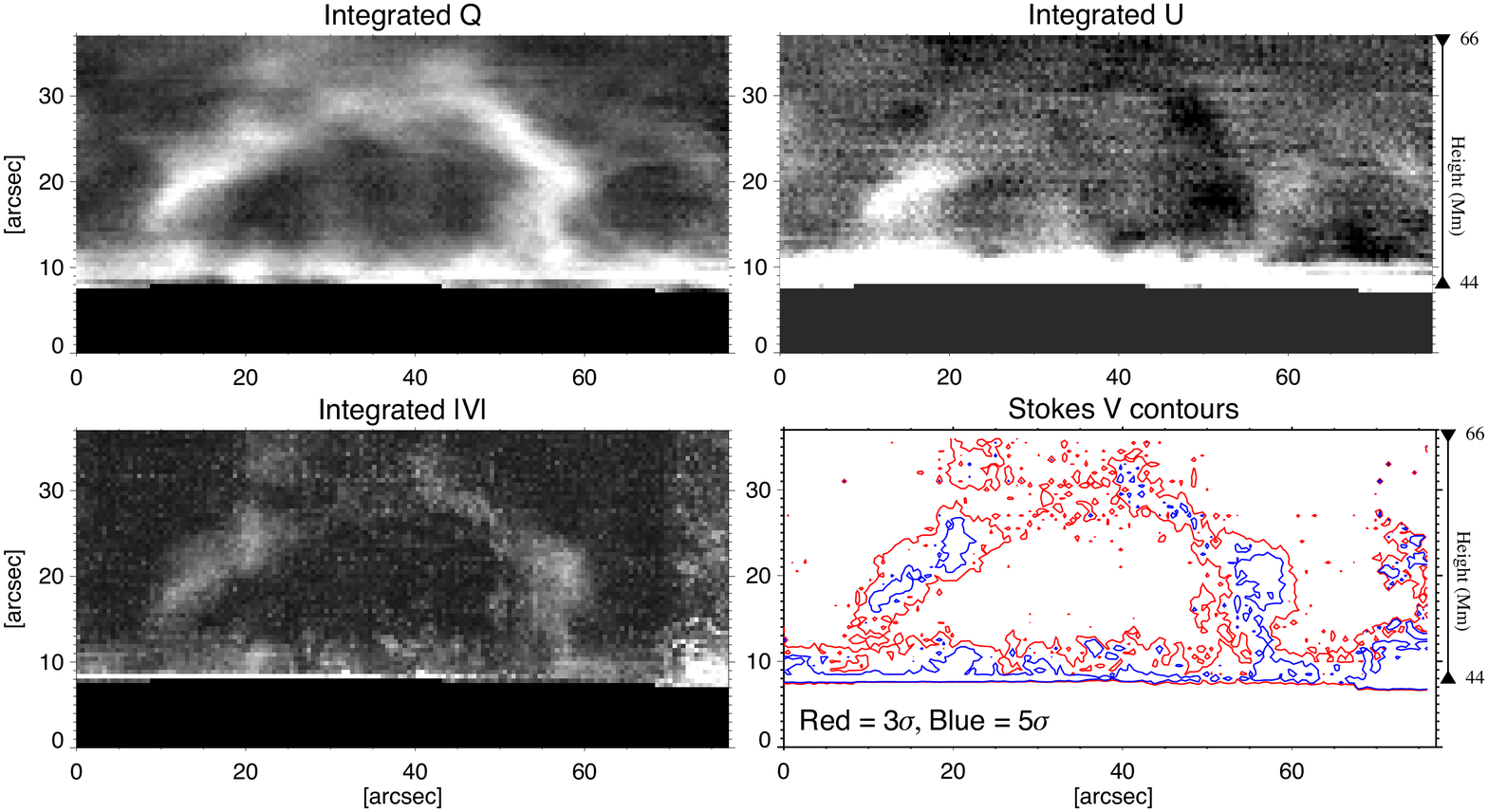}}
\end{center}
\caption{Stokes Q, U, and $|$V$|$ wavelength integrated maps calculated by integrating the observed Stokes profiles. The integral covered 21 wavelength samples centered on the position of maximum emission. The bottom-right panel shows contour plots representing the areas where the peak amplitude of the Stokes~$|$V$|$ signal surpasses three and five times the noise level $\sigma$. Note that $\sigma$ is pixel dependent since each Stokes parameters is normalized to its Stokes I maximum amplitude. As in Fig~\ref{fig1}, the bottom part represents the limb and the right axis the height in Mm.}
\label{fig5}
\end{figure*}

In the H$_\alpha$ image (bottom left panel), the prominence body is seen in emission outside the solar disk while the prominence feet are seen in absorption, hence darker that the surroundings, because they lie within the solar disk. The shape resembles the prominence as seen in the peak intensity of the \ion{He}{I}~1083.0~nm triplet (Fig.\ref{fig1}). The last two panels of Fig.~\ref{fig2} display STEREO \ion{Fe}{XII}~195~\AA\/ and \ion{He}{II}~304~\AA\/ observations. The prominence can be clearly seen in absorption (i.e., as a filament) what allows us to know its position on the solar disk and the angle $\theta$ between our line-of-sight (LOS) and the solar radius vector through the observed point (hereafter the local solar vertical). This light scattering angle is necessary to properly invert the Stokes profiles (see Sect.\ref{sec4}). In this case, $\theta\approx70\degr\/$, on average. The exact values of the $\theta$ angle at each pixel of the TIP-II slit is taken into account in the analysis of the profiles with HAZEL. We can also calculate the real height $h$ over the solar surface from the apparent height $h^\prime$ as  (see Fig.~3 from \citealt{2006ApJ...642..554M}):
\begin{equation}
 h=\frac{R_{\odot}+h^\prime}{\cos{(90\degr-\theta)}}-R_{\odot}. 
 \end{equation}
 In Fig.~\ref{fig3} we sketch the geometry of the prominence. In particular, we show a top view of the prominence using STEREO-B contour lines (right side). The filament has a length of about 138\arcsec\/ ($\sim$~100~Mm) and a width of 15\arcsec. The angle between the LOS and the long axis of the prominence is $\alpha = 90\degr+\beta$, where $\beta\sim17$\degr\/ is the angle that the prominence forms with the meridian, measured counterclockwise. In the right side of Fig.~\ref{fig3} we define the reference system with the Y and Z-axis contained in the sky-plane. Finally, SDO/HMI magnetograms provided us with a map of the photospheric magnetic flux. They show that in the right side (solar west) of the prominence positive polarity flux dominates the photosphere. This information will help us to determine the chirality of the filament, once we have inferred the magnetic field vector in the prominence body.

We classify the observed prominence as of quiescent type meaning that it is located outside active regions. Quiescent prominences are often characterized as sheets of plasma standing vertically above the PIL and showing prominence threads. When these threads are vertically oriented, quiescent prominences are often classified as of hedgerow type. The attained spatial resolution in TIP-II observations prevented us from resolving any prominence small-scale structures. However, we do see strands in the EUV images. At high latitudes, these show motions mainly parallel to the solar limb, similar to those described by \cite{2008ApJ...689L..73C}. At low latitudes, the motions seem to be perpendicular to the limb. These motions are typical of interactive hedgerow prominences \citep{1943ApJ....98....6P,1985SoPh..100..415H}. The long-term evolution of the prominence can be seen in an SDO/AIA \ion{Fe}{XIV}~211~\AA\/ movie available in online edition along with Fig 2. In the movie, the evolution of the fine scale structures can be very well appreciated. Interestingly, at 16:30 UT (5 hours after the TIP-II observation) and for no apparent reason, the prominence begins to rise and erupts (not shown in the movie). This prominence may be similar to the one observed by \cite{1983SoPh...89....3A} which also erupted shortly after the observations. The eruption is slow and last 15 hours until 21 May 7:00 UT. Remarkably, we have detected apparent spiraling motions in the feet of this prominence on 19 May 2011 using sit-and-stare slit observations and TIP-II \citep{2012ApJ...761L..25O}.

\section{Analysis of the polarization signals}
\label{sec3}

Individual \ion{He}{i}~1083.0~nm triplet emission profiles recorded in this 
quiescent prominence are shown in Fig.~\ref{fig4} (open dots). The Stokes I 
profiles are normalized to their maximum peak value (first column). They show 
the fine structure of the  \ion{He}{i}~1083.0~nm triplet, i.e., a weak blue 
component at 1082.9~nm ($^3$S$_1-^3$P$_0$) separated about 0.12~nm from the two 
blended components located at about 1083.03~nm ($^3$S$_1-^3$P$_1$ and 
$^3$S$_1-^3$P$_2$) that produce a stronger emission peak. The next columns 
represent the Stokes Q/I$_\mathrm{max}$, U/I$_\mathrm{max}$, and 
V/I$_\mathrm{max}$ signals, normalized to each of their Stokes I maximum 
value\footnote{In off limb observations, it is typical to normalize the 
polarization signals to their Stokes I peak value since there is no emission in 
the continuum.} (hereafter for simplicity Q/I, U/I, and V/I). The Stokes Q/I and 
U/I polarization signals are given after rotating the reference system so that 
the positive reference direction for Stokes Q is the parallel to the nearest 
limb. They show nonzero profiles with a prototypical Stokes I shape but with the 
blue component absent. This is the signature of scattering polarization, whose 
true physical origin is the presence of atomic level polarization. The mere 
presence of the Stokes U/I signal suggests the presence of a magnetic field 
inclined with respect to the local vertical direction, according to the Hanle 
effect theory. Note that, both, Stokes Q/I and U/I show polarization only in the 
red component of the triplet, as expected for the case of a prominence observed 
against the dark background of the sky 
\citep{2002Natur.415..403T,2007ApJ...655..642T}. 
 The blue component does not show any linear polarization signal in weakly magnetized optically
thin plasmas observed against the dark background of the sky, because its linear polarization can only be due to the selective absorption of polarization components caused by the presence of atomic polarization in the (metastable) lower level of the \ion{He}{i}~1083.0~nm triplet. Since
the upper level of the blue component ($^3$P$_0$) has $J = 0$, it cannot be polarized,
so that the emitted radiation has to be unpolarized. Only when dichroic effects
become important (in plasmas with a sizable optical depth or when observing plasma structures (e.g., quiescent filaments) against the bright background of the solar disk), the blue component may show
polarization. According to our observations, the prominence material we observed had to
present a quite small optical thickness (see Fig.~9), so that the
slab-model is suitable for the interpretation of the emission
profiles.

Finally, the Stokes V/I signals show the  typical circular polarization profiles 
dominated by the Zeeman effect, i.e., two lobes of opposite sign with and a 
zero-crossing point. Overall, Fig.~\ref{fig4} illustrates nicely the joint 
action of the Hanle and Zeeman effect in the \ion{He}{i}~1083.0~nm triplet. 
We display three different cases: (a) represents a pixel in  which the circular 
polarization signal is well above the noise level and shows the prototypical 
antisymmetric Stokes V profiles  dominated by the longitudinal Zeeman effect; 
(b) corresponds to a pixel with negligible circular polarization; and (c) shows 
a profile with net (wavelength-integrated) circular polarization in Stokes V/I. 
The physical origin of the net circular polarization in case (c) may be due to 
the presence of atomic orientation\footnote{It refers to the presence of 
population imbalances
and quantum coherence between the magnetic sub-levels of a given level. We refer 
the reader to \citep{2012ApJ...759...16M} and references therein.} in the 
energy levels and/or to the presence of correlated magnetic field and velocity 
gradients along the LOS \citep{1996SoPh..164..191L}. 

Figure \ref{fig5} displays Stokes Q, U, and $|$V$|$ wavelength integrated maps. The Stokes Q map closely resembles the Stokes I peak intensity map displayed in Fig.~\ref{fig1}. Here, the feet and the prominence body are clearly distinguishable form the background. In the case of Stokes U, the map shows clear signals in the prominence feet but not in the prominence body. Interestingly, the sign of Stokes U changes from positive (white) in the left feet to negative (black) in the right feet. There are some positive signals in the right feet as well. Thanks to the long exposure times per slit position, we could also detect clear circular polarization signals along the prominence. In this case, the  strongest Stokes V signals are concentrated in the feet of the prominence while in the body they are rather weak and contaminated by the noise. In the bottom right panel we display contours delimiting areas where Stokes V surpasses three and five times their intrinsic noise levels \footnote{Since the Stokes profiles are normalized to the peak amplitude of Stokes I, the noise level $\sigma$ (or the signal-to-noise ratio) differs from pixel to pixel, depending on the amplitude of the detected signal.} $\sigma$. Here, the fact that the Stokes V signals dominate in the feet is evident. As we explain in  Section 4.2, the mere detection of circular polarization is important in order to distinguish between the different magnetic field vector orientations compatible with the observed linear polarizations signals. Moreover, it helps to determine the strength of the magnetic field when the  \ion{He}{i}~1083.0~nm triplet is in the saturation regime ($B\gtrsim10$~G). 

\section{Interpretation of the polarization signals}
\label{sec4}

\subsection{Diagnostics of the \ion{He}{i}~1083.0~nm triplet}

The \ion{He}{i}~1083.0~nm triplet is suitable for the determination of the 
magnetic field vector in chromospheric and coronal structures. The main reason is that 
there are three physical processes able to generate and/or modify circular and 
linear polarization signals in the \ion{He}{i}~1083.0~nm triplet: the Zeeman 
effect, atomic polarization resulting from anisotropic radiation pumping, and 
the Hanle effect. All these effects can be studied and understood within the 
framework of the quantum theory of spectral line polarization 
\citep{2004ASSL..307.....L,2002Natur.415..403T,2007ApJ...655..642T} and provide 
a broad sensitivity to the magnetic field vector, from very weak fields to hecto- and kilo-Gauss fields. 
To interpret the observations we use the HAZEL code. This code is able to infer 
the magnetic field vector from the emergent Stokes profiles of the 
\ion{He}{i}~1083.0~nm triplet considering the joint action of atomic level 
polarization and the Hanle and Zeeman effects. The following assumptions made
by the HAZEL code are particularly fulfilled in solar prominences:

\begin{itemize}
\item We have chosen a simple radiative transfer scenario where the radiation is 
produced by a slab of constant physical properties. Neglecting or including the magneto-optical 
effect terms of the propagation matrix, the solution to the radiative transfer
equation has an analytic expression \citep{2008ApJ...683..542A,2005ApJ...619L.191T}. Under this approximation, the 
optical thickness of the slab, $\Delta\tau$, is a free parameter that can be inferred 
from the Stokes I profile at each point of the observed field of view. 

\item The slab atoms are illuminated from below by the (fixed and angle-dependent) photospheric solar 
continuum radiation tabulated by \cite{cox00}, producing population 
imbalances and quantum coherence in the levels of the \ion{He}{i} atoms. This produces
polarization in the emitted radiation. 

\item The atomic level polarization are calculated assuming complete frequency 
redistribution, which is a reliable approximation for modeling the observed polarization 
signatures in the \ion{He}{i}~1083.0~nm triplet \citep{2002NCimC..25..783T,2010mcia.conf..118T}.
\item We work in the collisionless regime, in which the atomic polarization is controlled
by radiative processes. No reliable estimations of the depolarizing collisional rates are
available for the \ion{He}{i} atom.
\end{itemize}

The inference strategy in HAZEL consists in comparing the observed Stokes profiles with synthetic 
guesses of the signals. Within HAZEL, the slab model is fully described 
using seven free parameters: the optical slab's thickness $\Delta\tau$ at the central wavelength of the 
red blended component, the line damping parameter $a$, the thermal velocity 
$\Delta\mathrm{v}_\mathrm{D}$, the bulk velocity of the plasma $v_\mathrm{LOS}$, 
and the strength $B$, inclination $\theta_\mathrm{B}$, and azimuth 
$\chi_\mathrm{B}$ of the magnetic field vector with respect to the solar 
vertical. To fully characterize the incoming radiation we also need to determine 
the height $h$ above the limb of each prominence point and the  $\theta$ angle, 
i.e., the angle that forms the LOS direction with the local vertical (see 
Fig.~\ref{fig3}). These two parameters fix the degree of anisotropy of the 
incident radiation field and are kept constant since we can determine them 
directly from the observations. It is important to properly determine the real 
height above the limb as well as the $\theta$ angle, to avoid imprecisions on 
the determination of the field vector orientation. In our case, these two 
parameters were determined using STEREO data. Finally, note that only pixels 
whose linear polarization signals exceeded five times the noise level are inverted in 
order to minimize the effect of noise on the inferences.

\subsection{Ambiguities}

\begin{figure}
\begin{center}
\resizebox{0.48\hsize}{!}{\includegraphics{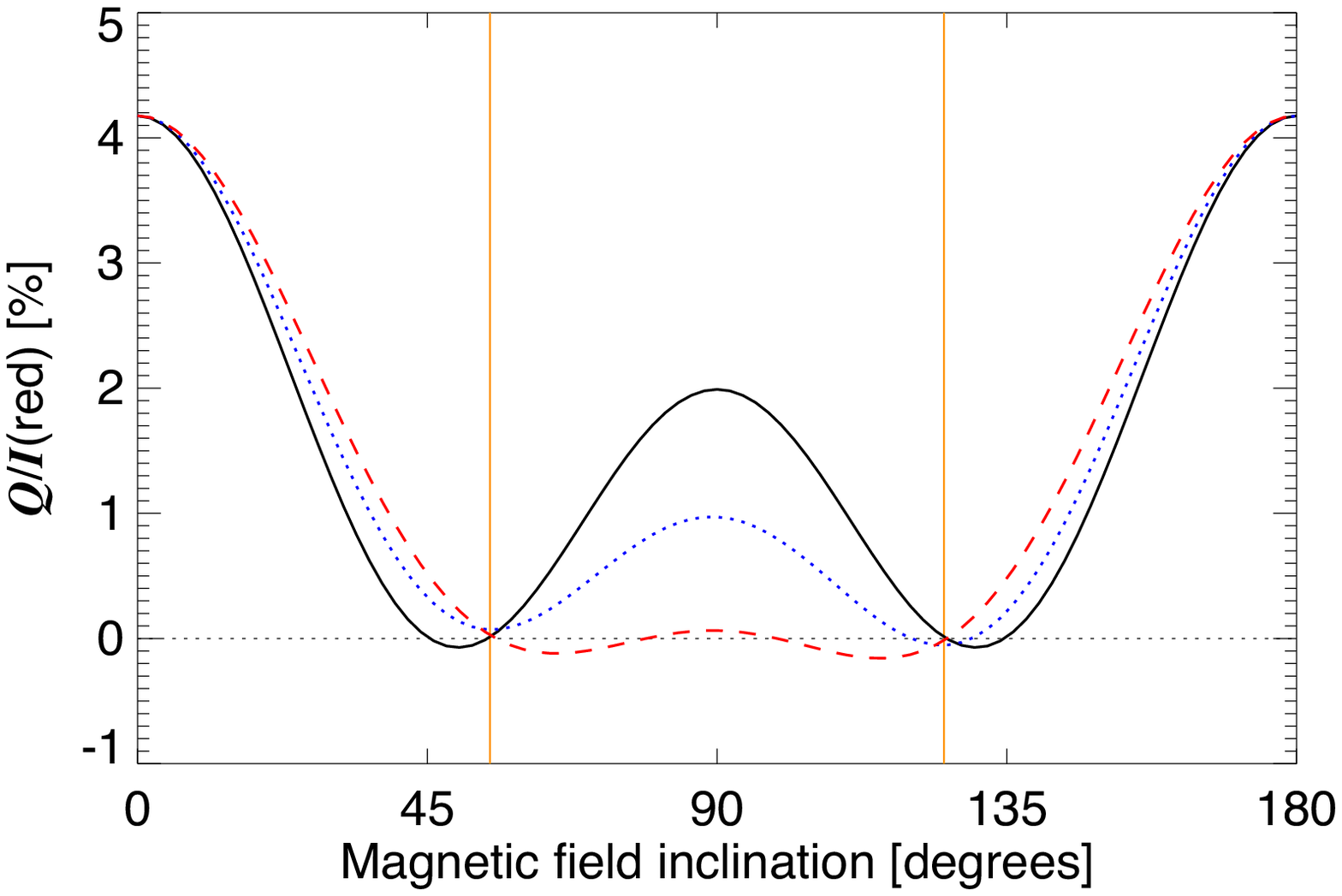}}
\resizebox{0.48\hsize}{!}{\includegraphics{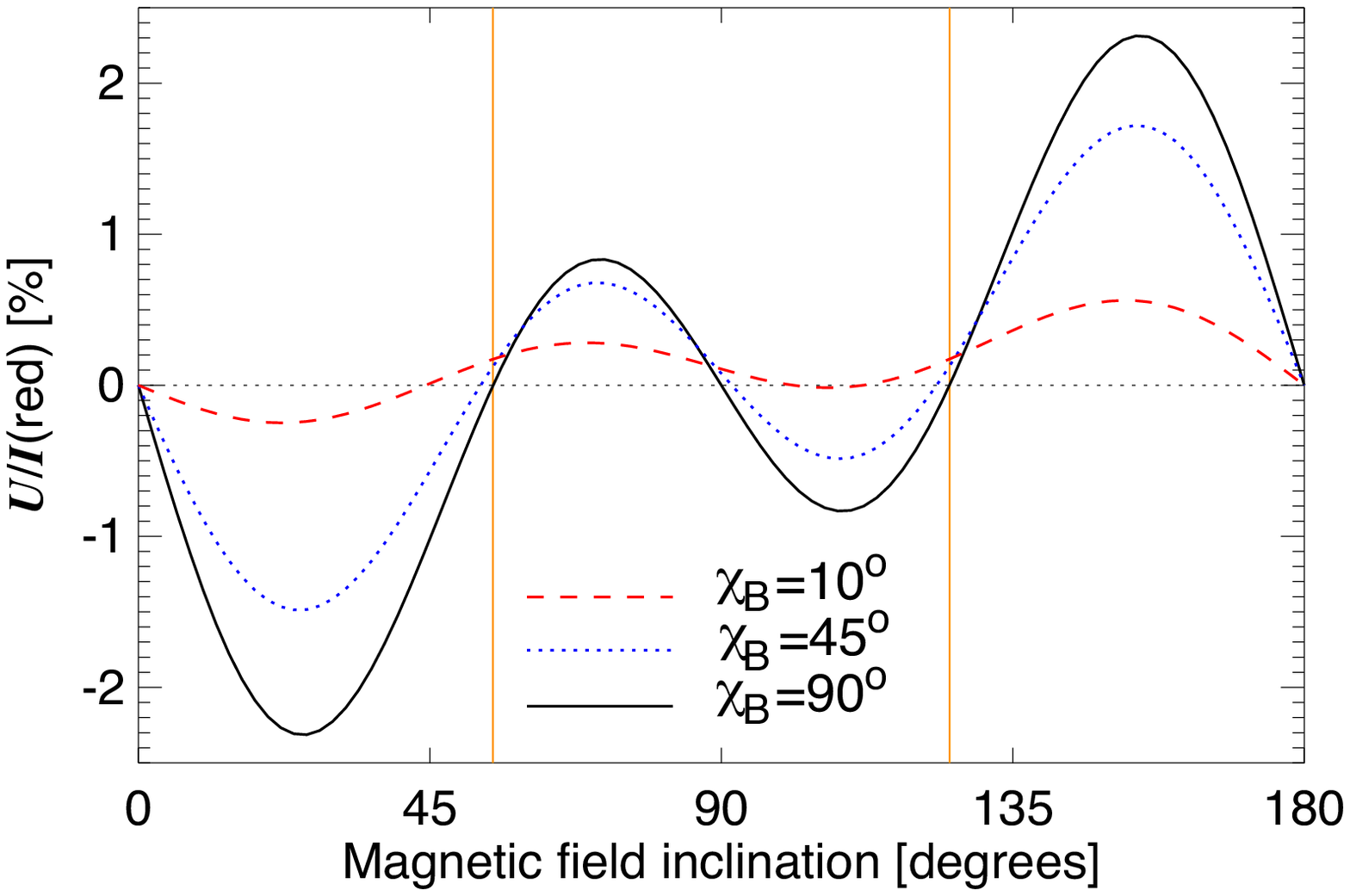}}
\resizebox{0.48\hsize}{!}{\includegraphics{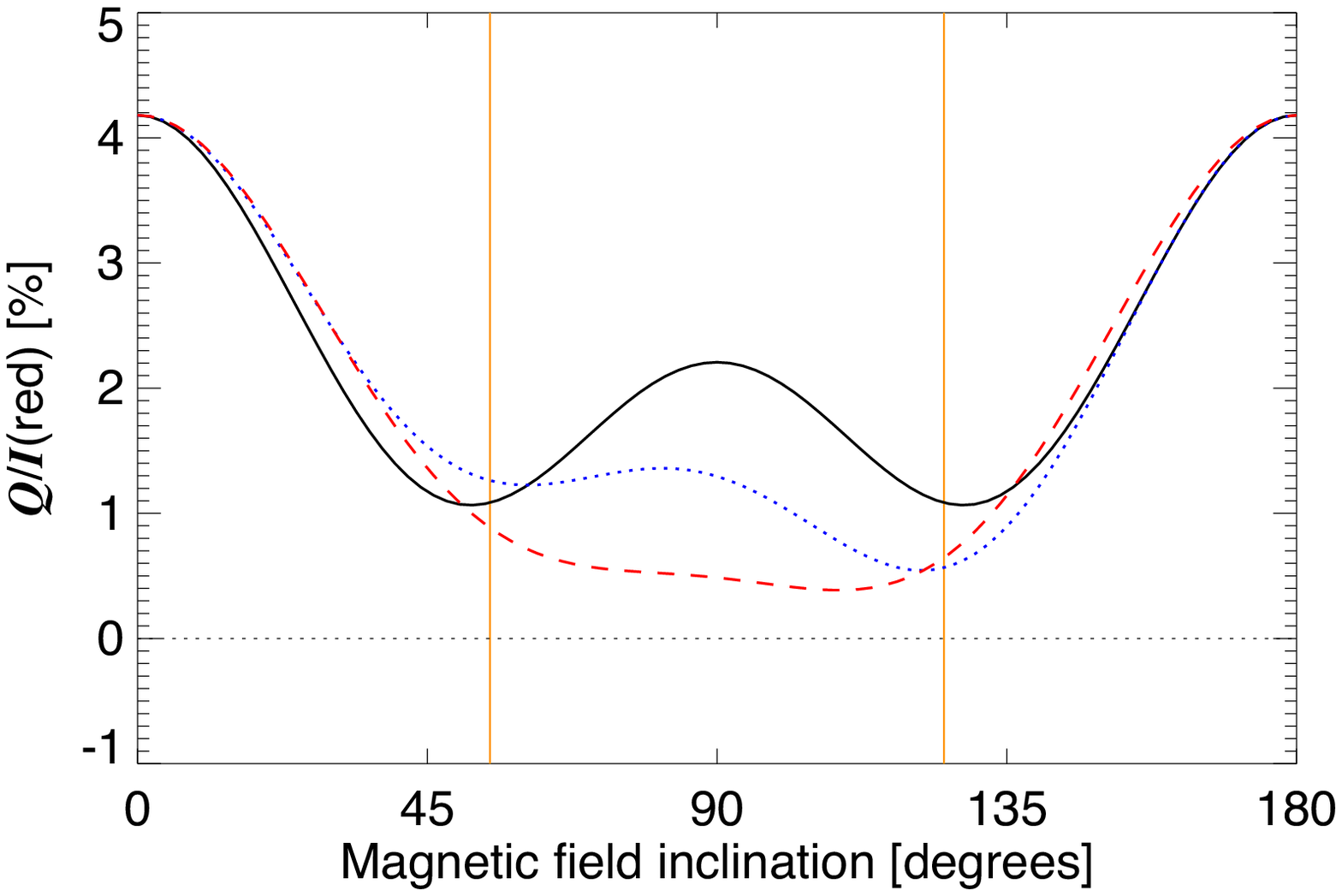}}
\resizebox{0.48\hsize}{!}{\includegraphics{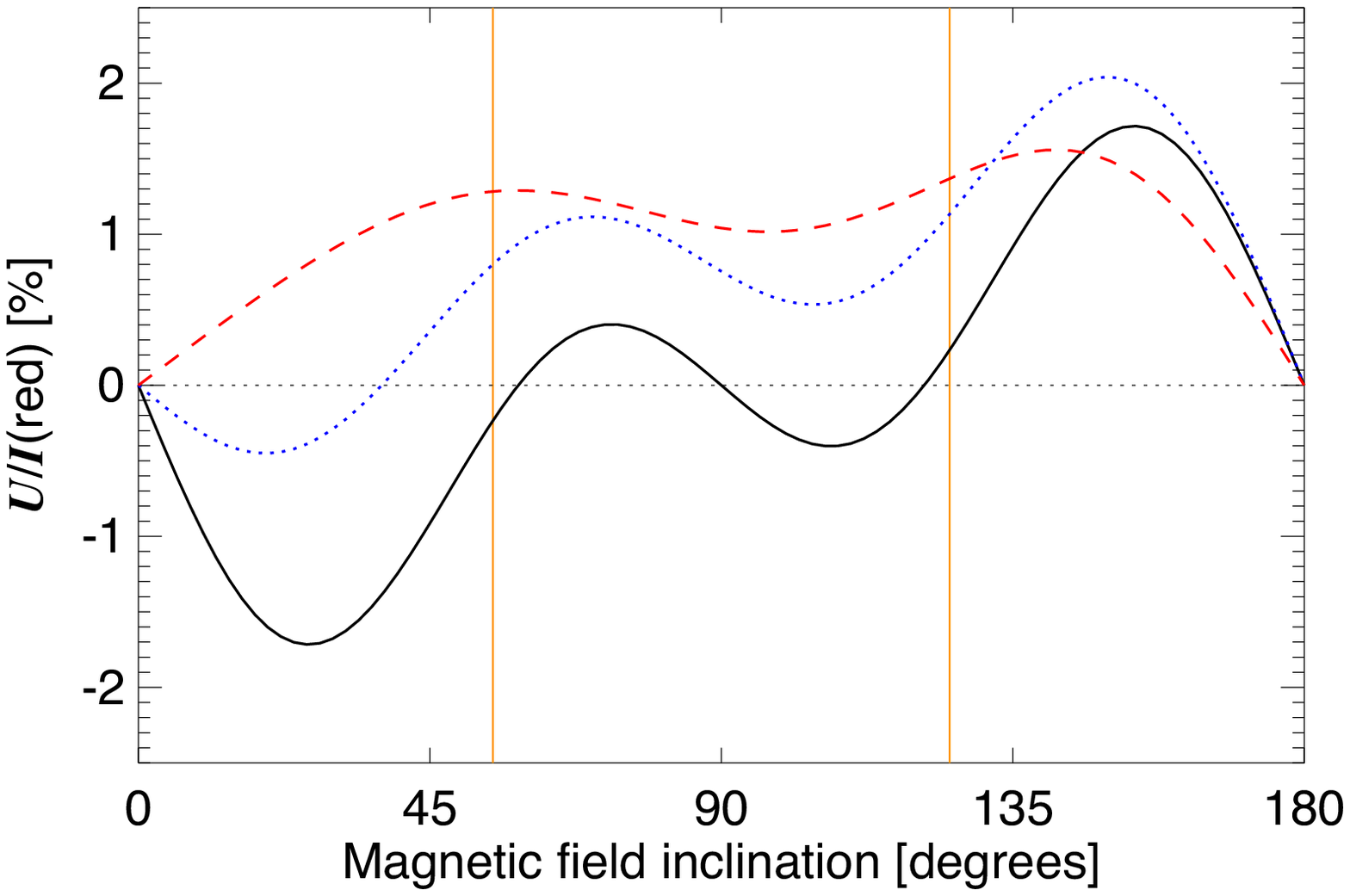}}
\end{center}
\caption{Variation of Stokes Q/I and U/I amplitudes of the \ion{He}{i}~1083.0~nm triplet red component with the magnetic field inclination for three different magnetic field azimuth values, in the 90\degr\/ scattering geometry. The top panels correspond to the saturation regime, $B > 10$~G, and the bottom panels to no saturation, $B = 1$~G. The rest of parameters for synthesizing the Stokes profiles were: $\Delta\tau=0.6$ at the central wavelength of the red blended component, $a=0.5$,  $\Delta\lambda_\mathrm{D}=7$~km~s$^{-1}$, and $h=10$\arcsec. The vertical lines are the Van Vleck angles at 54.74\degr\/ and 125.26\degr. Here, the positive reference direction for Stokes Q corresponds to $\chi_\mathrm{B}=$90\degr.}
\label{fig6}
\end{figure}

The behavior of the \ion{He}{i}~1083.0~nm triplet polarization signals depends in a very complicated manner on the orientation of the magnetic field with respect to the LOS and on the inclination of the magnetic field with respect to the local solar vertical. This dependence can be clearly seen in the Hanle saturation regime. In this case, the amplitude of the linear polarization signals is given by (Eq.~6 of \citealt{2010mcia.conf..118T})
\begin{equation}
\frac{Q}{I} \approx -\frac{3}{4\sqrt{2}}\sin^2\Phi_\mathrm{B} (3\cos^2\theta_\mathrm{B}-1)\mathcal{F},
\label{eqn1}
\end{equation} 
where $\Phi_\mathrm{B}$ is the angle between the magnetic field vector and the LOS, $\theta_\mathrm{B}$ is the inclination of the magnetic field vector with respect to the local solar vertical, and we take the Stokes Q reference direction the parallel to the projection of the magnetic field onto the plane perpendicular to the LOS (i.e., the reference direction for which Stokes U/I = 0). The quantity $\mathcal{F}$ is determined by solving, for the unmagnetized case,  the statistical equilibrium equations for the elements of the atomic density matrix. The $(3\cos^2\theta_\mathrm{B}-1)$ term tells us that Stokes Q/I is identically zero when $\theta_\mathrm{B}$ is 54.74\degr\/ and 125.26\degr. These are the so-called Van Vleck angles. Given the non-linear function of Eq.~\ref{eqn1}, we can expect that Stokes Q/I takes the same value for different field inclinations. This non-linear behavior gives rise to the so-called 90\degr\/ ambiguity of the Hanle effect. The 90\degr\/ ambiguity is associated with the Van Vleck angles because the magnetic field inclination of two 90\degr\/ ambiguous solutions lie at both sides of any of the Van Vleck angles. There can be particular orientations of the field vector where this ambiguity does not take place (e.g., \citealt{2006ApJ...642..554M}). Additional material about how this ambiguity works and how to deal with it can be found in \cite{2008ApJ...683..542A,2005ApJ...622.1265C,2009ApJ...703..114C,2006ApJ...642..554M}.

A simplified illustration of why the 90\degree\/ ambiguity appears can be found in Fig.~\ref{fig6}. The figure shows the dependence of the Stokes Q/I and U/I amplitudes, measured at the central wavelength of the red blended component of the \ion{He}{i}~1083.0~nm triplet, with the magnetic field inclination for different azimuth values. All calculations were done with the HAZEL code. The graph is similar to that of figure 9 in \cite{2008ApJ...683..542A} but for the 90\degr\/ scattering geometry. The top panels correspond to a magnetic field strength $B=11$~G, for which the \ion{He}{i}~1083.0~nm triplet is practically in the saturation regime of the upper-level Hanle effect. The bottom panels is for $B=1$~G. Note the non-linear dependence of the Stokes Q/I and Stokes U/I amplitude signals with the inclination angle. Because of this non-linear dependence, there may be particular magnetic field vector orientations giving rise to the same Stokes Q/I or U/I amplitude signals. For instance, a magnetic field configuration having $90^\circ<\theta_\mathrm{B}<125.26^\circ$ and $\theta_\mathrm{B} < 54.74^\circ$ can potentially give rise to the same Stokes Q/I and U/I amplitudes by appropriately modifying the field azimuth. 

In the Hanle saturation regime, the linear polarization signals depend only on the inclination of the magnetic field $\Phi_\mathrm{B}$ with respect to the LOS and on the inclination of the magnetic field $\theta_\mathrm{B}$ with respect to the local solar vertical (see Eq.~1). In the non saturated case the variation of Stokes Q/I and U/I amplitudes with $\theta_\mathrm{B}$  depends also on the field strength $B$. As a result, the Stokes Q/I signals may be always positive for any field inclination and azimuth. Stokes U/I can also be always positive when $\chi_\mathrm{B}=0^\circ$. Moreover, the Stokes Q/I and U/I amplitude signals are not symmetrical around $\theta_\mathrm{B}=90^\circ$ in the not saturated regime. This strong dependence on field strength, helps determining not only the orientation of the magnetic field vector from the analysis of the linear polarization signals, but also the field strength itself.
 
\begin{figure}
\begin{center}
\resizebox{0.8\hsize}{!}{\includegraphics{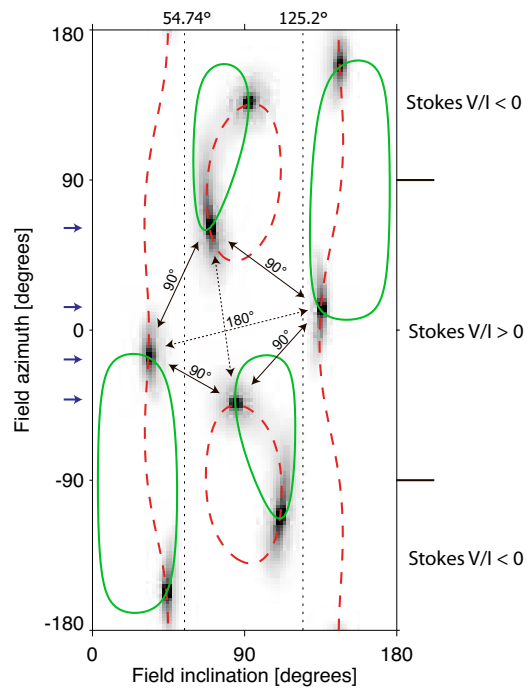}} 
\end{center}
\caption{Variation of the merit function with the inclination $\theta_\mathrm{B}$ and azimuth $\chi_\mathrm{B}$ of the field vector. The minima are located at eight different positions. Four of them correspond to a field vector pointing to the observer, i.e., positive Stokes V signals. These four solutions are connected through the 90\degr\/ ambiguity and the 180\degr\/ ambiguity of the Hanle effect (see arrows). The vertical dotted lines correspond to the Van Vleck angles 54.74\degree\/ and 125.26\degree. The merit function here has been calculated using the parameters derived from the inversion of profile (a) in Fig.~\ref{fig4}: $B = 17$~G, $\Delta\tau=1.13$, $\theta = 71.4$\degr, $h=$~61\farcs8, $\Delta\lambda_\mathrm{D}=6$~km~s$^{-1}$, and $a=0.33$. The contours delimit connected regions by the same value of Stokes Q/I (dashed) and U/I (solid).}
\label{fig7}
\end{figure}

Besides the 90\degr\/ ambiguity, there is another ambiguity called the 180\degr\/  ambiguity of the Hanle effect, similar to the well-known azimuth ambiguity of the Zeeman effect. In this case, the linear polarization signals do not change when we rotate the field vector 180\degr\/ around the LOS axis (this is a rotation of the magnetic field in the plane perpendicular to the LOS). In the case of 90\degr\/ scattering geometry, profiles differing by 180\degr\/ in the inclination $\theta^*_\mathrm{B}=180^\circ-\theta_\mathrm{B}$ (in the plane perpendicular to the LOS) give rise to the same linear polarization profiles but with an azimuth $\chi^*_\mathrm{B}=-\chi_\mathrm{B}$. 

In summary, the determination of the orientation of the field vector from the linear polarization profiles of the \ion{He}{1}~1083.0~nm triplet is affected by two ambiguities: the 180\degr\/ azimuth ambiguity and the 90\degr\/ ambiguity. There is another pseudo-ambiguity associated with Stokes V/I. When the signal-to-noise-ratio is not high enough so as to detect the circular polarization signal, we cannot determine if the field vector is pointing to the observer or away from the observer. In 90\degr\/ scattering geometry it would correspond to an ambiguity in the azimuth as $\chi^\prime_\mathrm{B} = 180^\circ - \chi_\mathrm{B}$, that corresponds to specular reflections in the plane perpendicular to the LOS. Thus, in total we can have potentially eight possible solutions compatible with a single observation. 

As in Fig.~13 of \cite{2008ApJ...683..542A}, we show in Fig.~\ref{fig7} an example of how the merit function, calculated as $\chi^2 = \sum_{i=1}^2[S^\mathrm{syn}_i(\lambda_\mathrm{red})-S^\mathrm{obs}_i(\lambda_\mathrm{red})]^2$, varies with the inclination $\theta_\mathrm{B}$ and azimuth $\chi_\mathrm{B}$ of the field vector. In the previous expression, $i$ represents Stokes Q/I and U/I and $\lambda_\mathrm{red}$ is the wavelength position of the \ion{He}{1}~1083.0~nm red component. The figure shows eight local minima (dark places). Four of them are incompatible with the observations because they require Stokes V/I profiles of negative sign \footnote{Here, the sign of Stokes V is defined as the sign of the Stokes V/I blue lobe amplitude} (field vector pointing away from the LOS) and we observe a Stokes V/I with positive sign. The other four solutions (marked with arrows in the left axis) are connected through the 90\degr\/ ambiguity and the 180\degr\/ ambiguity as the solid and dotted arrow lines show. The solid and dashed contour lines represent isolines along which the merit function has  constant Stokes Q/I and U/I values. Note that they cross where the merit function has a local minima. 

In practice, to determine all possible solutions we proceeded as follows: first, we checked the sign of Stokes V/I to delimit the range of possible azimuths. In our prominence, Stokes V/I was always of positive sign, thus $-90^\circ<\chi_\mathrm{B}<90^\circ$, which means that the LOS magnetic field vector component is always pointing toward the observer. There are four potential solutions within the parameter range $-90^\circ<\chi_\mathrm{B}<90^\circ$ and $0^\circ<\theta_\mathrm{B}<180^\circ$. Therefore, to explore all possible solutions we perform six different inversions: three inversion runs allowing the azimuth $\chi_\mathrm{B}$ to vary between $-90^\circ$ and $0^\circ$ and the inclination to vary within the ranges $0^\circ<\theta_\mathrm{B}<54.74^\circ$, $54.74^\circ<\theta_\mathrm{B}<125.26^\circ$, and $125.26^\circ<\theta_\mathrm{B}<180^\circ$, and other three runs for $0^\circ<\chi_\mathrm{B}<90^\circ$ and the same inclination ranges. Only four of these inversion will provide a magnetic field vector configuration whose emergent Stokes profiles are compatible with the observations. This approach is rather rough although it allows to explore the complete parameter range. The present version of the HAZEL code does this work automatically, using as a basis the analytical expressions in the saturation regime. Note also that HAZEL provides a discrete solution for each pixel. The above strategy allows us of impose a continuity condition since for each of the inversion runs the solutions will not jump from one possible solution to another.

\section{Inversion results}

\begin{figure*}[!t]
\begin{center}
\resizebox{0.9\hsize}{!}{\includegraphics{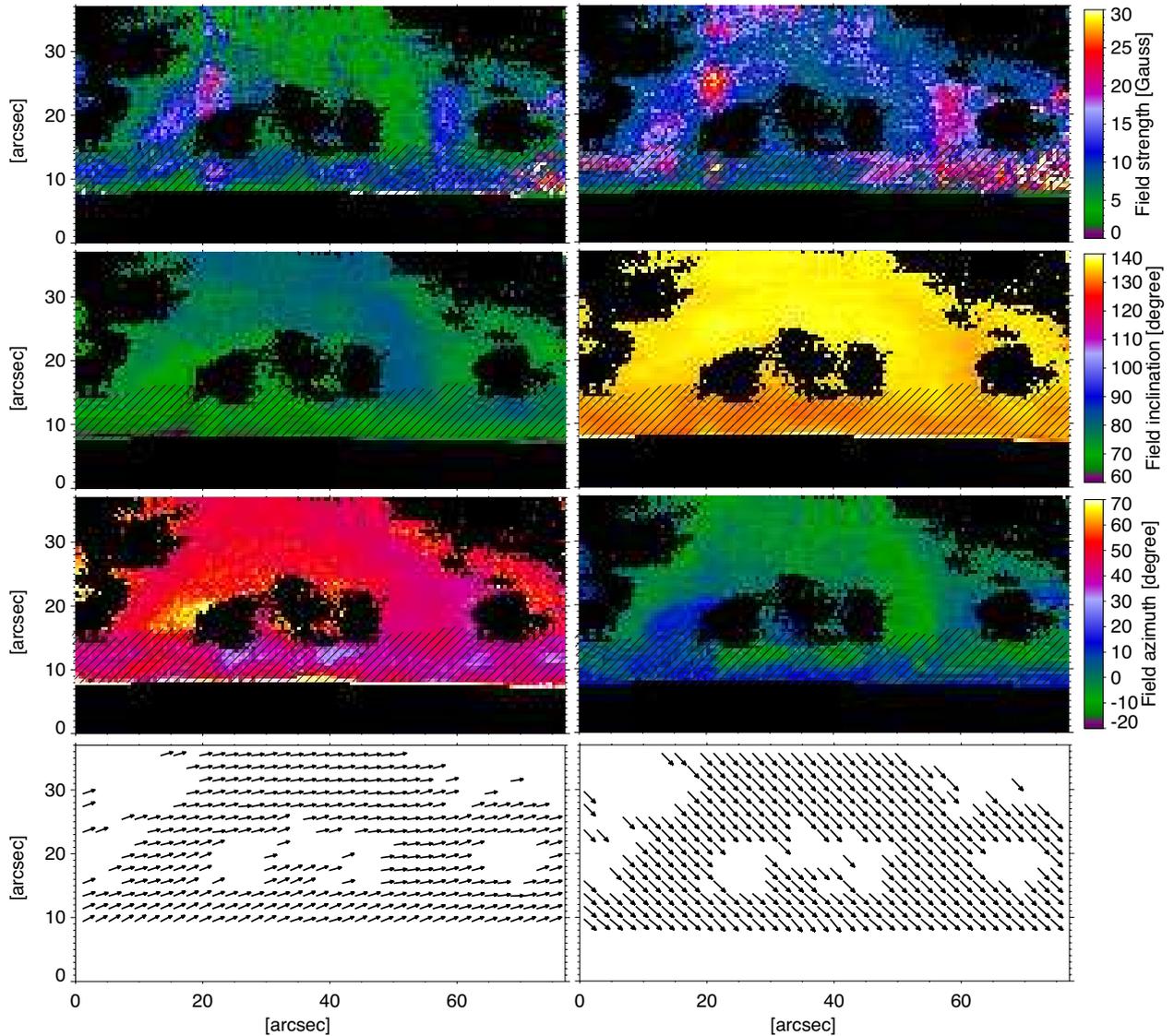}}
\end{center}
\caption{Field strength, inclination, and azimuth maps resulting from the inversion of the observed Stokes profiles with the HAZEL code. Each column represents one of the two 90\degr\/ ambiguous solutions, quasi-horizontal (left) and quasi-vertical (right). As in Figure~\ref{fig1}, the black bottom region represent the solar disk. The rest of dark areas correspond to pixels whose Stokes Q/I and U/I signals did not exceed 5 times their corresponding noise levels.}
\label{fig8}
\end{figure*}

\begin{figure}[!t]
\begin{center}
\resizebox{1.2\hsize}{!}{\includegraphics{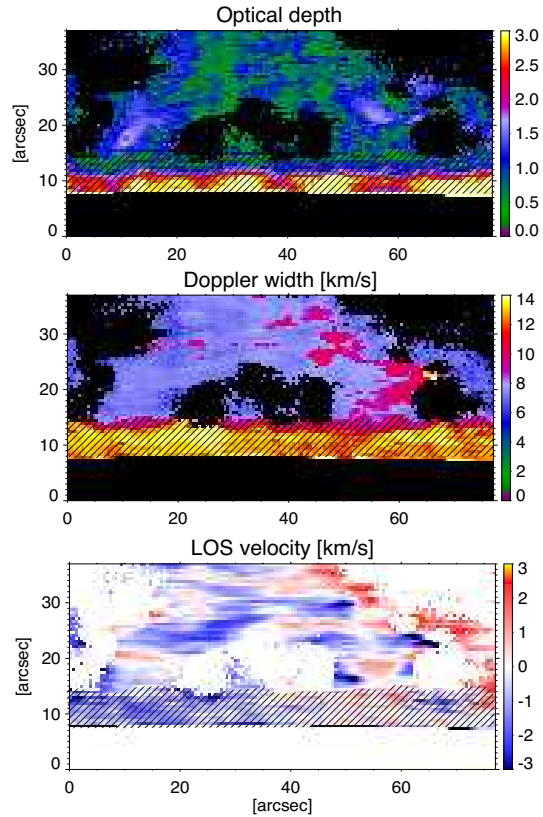}}
\end{center}
\caption{Optical depth, Doppler width, and line-of-sight velocity inferred with the HAZEL inversion code. The shaded area correspond to solar spicules. As in Fig.~\ref{fig7}, the line-shaded areas correspond to the spicules at about 8\arcsec--14\arcsec, excluded in our analysis.}
\label{fig9}
\end{figure}

Figure \ref{fig4} shows observed (open dots) and best-fit (solid) Stokes profiles of the \ion{He}{i}~1083.0~nm triplet corresponding to three different locations on the prominence. The sub-panels show the difference between the best-fit and  the observed profiles. Case (a) shows a profile where the linear and circular polarization signals are prominent above the noise level. The fits are good in Stokes Q/I, U/I, and V/I and the residuals are very small. Profiles in case (b) correspond to a point where the Stokes U/I and V/I signals are barely above the noise level. The fit in Stokes Q/I is good. Case (c) represents an infrequent pixel where the Stokes V/I signal shows net circular polarization. Again, the fits are good except for Stokes V/I. The version of HAZEL we applied neglects atomic orientation and possible correlations between the magnetic field and the velocity gradients along the LOS\footnote{Given that the fractional number of pixels showing anomalous Stokes V profiles is very small, we do not find it justified to increase the complexity of our model assumptions.}. Note that cases (a) and (c) are in the Hanle saturation regime, $B>10$~G. In the same panels we show both, the quasi-vertical and quasi-horizontal solutions, i.e., one solution and its corresponding 90\degr\/ ambiguous solution. Pay attention to the difference between the two fits: it is negligible and at the noise level (see horizontal dashed lines) which means that for two different magnetic field orientations the Stokes Q/I, U/I, and V/I profiles are indistinguishable. Two compatible solutions are found not only for two different magnetic field orientations but also for two different field strengths as it can be seen in the Stokes U/I legend. Only in case (c) the Stokes V/I fits differ, both of them fit the profile within the noise level of the data. 

The field strength, inclination, and azimuth maps of the prominence obtained from the analysis of the Stokes profiles with HAZEL and corresponding to two compatible solutions via the 90\degr\/ ambiguity are displayed in Fig.~\ref{fig8}. The left panels correspond to a solution where the inclination of the field vector is almost perpendicular to the local vertical (quasi-horizontal solution) and the right panels to a solution where the field vector is more vertical than horizontal (quasi-vertical solution). The analysis depicted an average field strength of about 7~G and 12~G for the quasi-horizontal and quasi-vertical solutions, respectively. These values are around the Hanle saturation regime which emphasizes the importance of detecting Stokes V in order to better constrain the strength of the inferred magnetic field. The maps show that the field strength varies smoothly along the prominence. Only in the left part (feet) of the prominence there is a region where the field is stronger than the average, with field strength values up to 25~Gauss for the quasi-horizontal case and 30~Gauss for the quasi-vertical one. A slight increase of the field strength is also noticeable in the right feet. Since for fields stronger than 10~G it is only possible to determine the orientation of the field vector from the Stokes Q/I and U/I profiles, it is important to detect Stokes V/I to help determine the field strength. In this case, the effective exposure time was high enough to measure significant Stokes Q/I, U/I, and V/I signals above the noise level (see Figure~\ref{fig5}).

In the quasi-horizontal solution, the inclination of the magnetic field vector, measured from the local vertical, varies from about 70$^\circ$ at the left part of the prominence to $\sim$~90$^\circ$ at the right side. The field inclination is about 135$^\circ$ for the quasi-vertical solution. Regarding the magnetic field azimuth, which is measured counterclockwise from the LOS direction, it is about 49$^\circ$ for the quasi-horizontal case and it varies between 10$^\circ$ and $-$10$^\circ$ for the quasi-vertical case. We can estimate the angle between the prominence long axis and the magnetic field vector as $\chi^\dag=\alpha - \chi_\mathrm{B}$ measured clockwise from the PIL, with $\alpha=107^\circ$ the angle between the LOS and the PIL. In this case $\chi^\dag$ is about 58$^\circ$ for the quasi-horizontal solution and $108^\circ$ for the quasi-vertical solution. The latter implies that the field vector is almost perpendicular to the filament PIL, which we may take as an extra reason to pot for the quasi-horizontal solution.

\begin{table}
\caption{Summary of results}              
\label{table:1}      
\centering                                      
\begin{tabular}{ccccc}          
\hline\hline                       
Quasi- & $B$ [G] & $\theta_\mathrm{B}$ [\degr]  & $\chi_\mathrm{B}$ [\degr] & $\chi^\dag$  [\degr]\\ 
\hline                                  
-horizontal (Dextral) & 6.9  & 77.0  &  49.0  & 58 \\   
-vertical (Sinistral) & 11.7   & 136.0 & -1.5 & 108 \\   
-horizontal (Sinistral -- 180\degr)   & 6.4 & 77.0 & -49.0 & 156 \\   
-vertical (Sinistral -- 180\degr)  & 7.8 & 33.8 & 1.2 & 106\\   
\hline                                  
\end{tabular}
\end{table}

The magnetic field vector configurations we show in Figure~\ref{fig8} correspond to two  possible solutions, related by the 90$^\circ$ ambiguity. All compatible solutions, including those resulting from the 180$^\circ$ ambiguity of the Hanle effect, are listed in Table 1. The table gives the average values of the inferred magnetic field strength $B$, field inclination $\theta_\mathrm{B}$, field azimuth $\chi_\mathrm{B}$, and the angle between the magnetic field vector and the PIL $\chi^\dag$. The table also specifies the chirality of the prominence, i.e., whether the solution is of dextral or sinistral chirality. The chirality can be determined using the information about the sign of the photospheric magnetic field which is positive (negative) at the right (left) side of the prominence. Dextral chirality corresponds to angles between $0^\circ<\chi^\dag<90^\circ$ and sinistral chirality to angles between $90^\circ<\chi^\dag<180^\circ$ \citep{1998ASPC..150..419M}. Since $\chi^\dag<180^\circ$ in all cases, the analyzed prominence is of inverse polarity, i.e., the direction of the prominence magnetic field vector is opposite to the direction expected for a potential field anchored in the photosphere.

In Fig.~\ref{fig9} we show the optical depth and the Doppler width resulting from the inversion of the profiles. The mean optical depth is 0.84 with slightly larger mean values at the prominence feet than at the prominence body. It increases significantly in the spicules. The Doppler width lies between 6~km~s~$^{-1}$ and 8~km~s~$^{-1}$ with a mean value of 6.4~km~s~$^{-1}$. In the same figure we display the inferred LOS velocity. It fluctuates between $\pm~3$~km~s~$^{-1}$. These values are in line with recent Doppler shifts measurements \citep{2010A&A...514A..68S}. 
 
 \begin{figure*}[!t]
\begin{center}
\resizebox{\hsize}{!}{\includegraphics{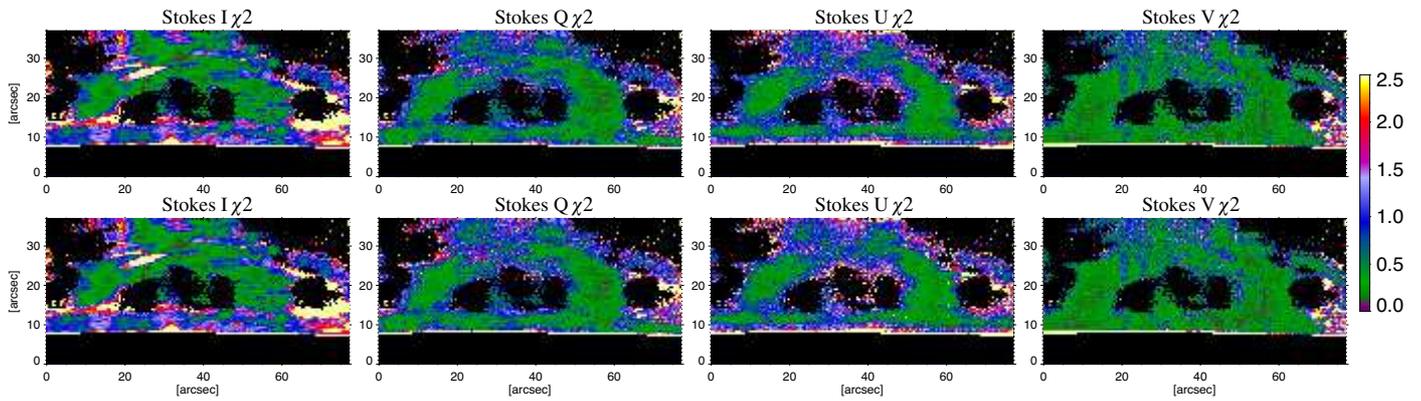}}
\end{center}
\caption{$\chi^2$ values resulting from the inversion of the data. The top panels correspond to one of the quasi-horizontal solutions and the bottom panels to one of the quasi-vertical solutions. As in Figure~\ref{fig1}, the black bottom region represent the solar disk. The rest of dark areas correspond to pixels whose Stokes Q/I or U/I signals did not exceed 5 times their corresponding noise levels.}
\label{fig10}
\end{figure*}

 Because of the complexity of the \ion{He}{i}~1083.0~nm triplet model and the 
inversion algorithm, which is based in global optimization methods, it is 
difficult to give a meaningful estimation of the statistical errors in the inversion results. In 
principle, we could use the last iteration of the inversion algorithm (based in a 
standard Levenberg-Marquardt technique) and estimate the covariance matrix. 
However, the obtained errors would depend on the exact definition of the merit 
function itself and would also neglect the effect of degeneracies. Thus, we consider that
the way to proceed in the future is to use a fully Bayesian approach. Unfortunately,
it is not yet implemented in HAZEL due to computational constraints.

However, with the merit 
function we can provide some information about the goodness of the fit. In Fig.~\ref{fig10} we provide the reduced $\chi^2$ values resulting from the 
inversion of the data \citep[see][]{2008ApJ...683..542A}. The top and bottom panels stand for the $\chi^2$ of the 
four Stokes parameters and for the quasi-horizontal and quasi-vertical 
solutions, respectively. First, note that the $\chi^2$ maps are very similar for 
the two solutions, which means that, from an statistical point of view, both 
solutions are equally probable. If we focus on individual panels, it can be seen 
how the $\chi^2$ slightly increases at the boundaries between the prominence and 
the background, where the signal-to-noise ratio is smaller. There are localized 
areas where the $\chi^2$ increases. For instance, in Stokes I, around 
x=20\arcsec and y=25\arcsec, there is a local enhancement of the $\chi^2$ 
values. This case is due to the presence of a second component in the Stokes 
I profiles. Overall, the figure shows that the fits are good for most of the 
pixels. 

\section{Summary and conclusions}
\label{sec5}

In this paper we have shown that spectropolarimetric observations in the \ion{He}{i}~1083.0~nm triplet are very useful  to determine the strength and orientation of the magnetic field in solar prominences. In particular, we have determined the magnetic field vector in a quiescent solar prominence, using observations of the \ion{He}{i}~1083.0~nm triplet taken with the TIP-II instrument installed at the VTT. Even though the integration time per slit position was of the order of one minute and the total time to scan the prominence took $\sim$~1.5 hour, it is possible to acquire data under stable observing conditions with TIP-II and still maintaining a moderate spatial resolution of 1\arcsec--1\farcs5. These integration times are necessary to increase the signal-to-the-noise ratio and detect both, circular and linear polarization signals in the prominence. These signals are often buried in the noise in ``standard'' TIP-II observations. The detection of the linear polarization signals allows us to determine the orientation of the field vector while Stokes V is crucial to fix the field strength.

To infer the field vector we have employed the HAZEL inversion code, which includes all necessary physics for interpreting the Stokes I, Q, U, and V profiles. We have shown that the use of context data, such as that provided by STEREO and SDO, may be crucial for setting up the scattering problem. In our case, STEREO allowed us to determine the prominence heights, its position on the solar disk (viewing angle with respect to the solar vertical), and the orientation of the PIL with respect to the solar limb (or to the LOS). 

The \ion{He}{i}~1083.0~nm triplet suffers from two ambiguities: the 90\degr\/ ambiguity  and the 180\degr\/ ambiguity of the Hanle effect. We have shown that the observed profiles themselves do not encode sufficient information to solve any of these ambiguities in 90\degr\/ scattering geometry (see Fig.~3). Fortunately, from theoretical arguments we can discard two of the solutions. For instance, we can assume that the magnetic field vector belongs to a weakly twisted flux rope or to a sheared arcade which also contains weakly twisted field lines. In this case, the prominence material would be located in dipped magnetic field lines. In these two models the component of the magnetic field vector along the prominence main axis dominates. Thus, the quasi-vertical solutions, which suggest that the magnetic field vector is almost perpendicular to the prominence main axis can be discarded. The quasi-vertical solution would also require a extremely highly twisted structure, which is rather improbable in prominences \citep{1989ApJ...343..971V}. Thus, there are two possible solutions only where the magnetic field vector is highly inclined, the quasi-horizontal solutions. These two solutions are connected through the 180\degr\/ ambiguity of the Hanle effect, i.e., they differ by a 180\degr\/ rotation in the plane perpendicular to the LOS. 

The quasi-horizontal solutions confirm previous findings about the average field strength in quiescent prominences, being about 7~G, in average. An interesting result is that the field strength seems to be more intense at the prominence feet, reaching values up to 30~G and coinciding with areas where the opacity increases. If the dense plasma is truly suspended in magnetic dips, the existing correlation between the opacity and the field strength may provide additional information to understand current physical mechanisms for suspending the prominence material. Interestingly we do not detect abrupt changes in the prominence field strength contrary to the results of \cite{2003ApJ...598L..67C}. 

Our results for the orientation of the field vector with respect to the solar surface slightly deviate from previous measurements. In particular, we found that the field vector is about 77\degr\/ inclined with respect to the solar vertical. This result is between the values reported by e.g., \cite{1989ASSL..150...77L} and \cite{1994SoPh..154..231B}, with inclinations of about 60\degr\/ from the local vertical, and those reported by \cite{2003ApJ...598L..67C}, mostly horizontal fields. 

Regarding the orientation of the field with respect the prominence main axis, we found it to be $\sim$~58\degr\/ in one case (dextral chirality) and $\sim$~156\degr\/ in the corresponding 180\degr\/ ambiguous solution (sinistral chirality). The first one differs from previous findings. For instance, \cite{1989ASSL..150...77L}, \cite{1994SoPh..154..231B}, and \cite{2003ApJ...598L..67C} report angles below 30\degr\/. In contrast, the second solution, $\chi^\dag=156^\circ$, implies an acute angle of $24^\circ$, in line with previous measurements. In practice, we cannot distinguish between the two of them. We point out that the sinistral chirality case is the most probable solution  for southern prominences \citep{martin}, although it may also be possible that the twisting is related with the fact that this prominence was ejected few hours after the observations. In this case, the more twisted case (dextral) would be the ``true'' magnetic configuration.

\begin{acknowledgements}
We thank  M.\ Collados and M.\ J.\ Mart\'inez Gonz\'alez for their invaluable help in the TIP-II data reduction. Financial support by the Spanish Ministry of Economy and Competitiveness and the European FEDER Fund through project AYA2010-18029 (Solar Magnetism and Astrophysical Spectropolarimetry) is gratefully acknowledged. AAR also acknowledges financial support through the Ram\'on y Cajal fellowship.
\end{acknowledgements}

\end{document}